\documentclass[12pt]{article}
\pdfoutput=1

\usepackage[utf8]{inputenc}
\usepackage[margin=1in]{geometry}
\usepackage[tbtags]{amsmath} 				% split tags on lower right
\usepackage{amssymb, mathrsfs}			% for math script fonts
\usepackage{physics}
\usepackage{graphicx}			% for images
\usepackage{tensor}				% for tensors
\usepackage[makeroom]{cancel}	% for cancel
\usepackage{bm}								% for bold Greek
\usepackage[
      colorlinks=true,
      linkcolor=blue,
      urlcolor=blue,
      filecolor=black,
      citecolor=red,
      pdfstartview=FitV,
      pdftitle={},
      pdfauthor={Michael Gutperle, Kevin Chen},
      bookmarksopen=true
      ]{hyperref}

\usepackage{graphicx,calc}
\newlength\myheight
\newlength\mydepth
\settototalheight\myheight{Xygp}
\DeclareMathOperator{\diag}{diag}

\DeclareMathOperator{\AdS}{AdS}
\DeclareMathOperator{\SO}{SO}
\DeclareMathOperator{\SU}{SU}

\newcommand{\cA}{\mathcal A}
\newcommand{\cB}{\mathcal B}

\newcommand{\cK}{\mathcal K}

\newcommand{\cP}{\mathcal P}
\newcommand{\cN}{\mathcal N}
\newcommand{\cM}{\mathcal M}
\newcommand{\cO}{\mathcal O}
\newcommand{\cV}{\mathcal V}
\newcommand{\cQ}{\mathcal Q}
\newcommand{\cL}{\mathcal L}

\def\bi{\bar\imath}
\def\bj{\bar\jmath}

\usepackage{sectsty}
\sectionfont{\large}

\renewcommand{\thefootnote}{\fnsymbol{footnote}}
\renewcommand{\thanks}[1]{\footnote{#1}}
\newcommand{\starttext}{
\setcounter{footnote}{0}
\renewcommand{\thefootnote}{\arabic{footnote}}}
\renewcommand\({\begin{equation}}		% quick macro for equation with numbers
\renewcommand\){\end{equation}}
\renewcommand{\epsilon}{\varepsilon}	% use varepsilon
\usepackage{dsfont}

\numberwithin{equation}{section} 		% number equations by section

\numberwithin{equation}{section}

\long\def\symbolfootnote[#1]#2{\begingroup%
\def\thefootnote{\fnsymbol{footnote}}\footnote[#1]{#2}\endgroup}

\begin{document}
\setlength{\baselineskip}{16pt}

\starttext
\setcounter{footnote}{0}

\begin{flushright}
\today
\end{flushright}

\bigskip

\begin{center}

{\Large \bf  Janus solutions in three-dimensional $ {\cal N}=8$ gauged supergravity}

\vskip 0.4in

{\large  Kevin Chen   and Michael Gutperle }

\vskip 0.2in

{\sl Mani L. Bhaumik Institute for Theoretical Physics}\\
{\sl Department of Physics and Astronomy }\\
{\sl University of California, Los Angeles, CA 90095, USA}

\end{center}
 
 \bigskip
 
\begin{abstract}
\setlength{\baselineskip}{16pt}

 Janus solutions  are constructed in $d=3$, ${\cal N}=8$ gauged supergravity. We find explicit half-BPS solutions where  two scalars in the  $\SO(8,1)/\SO(8)$ coset have a nontrivial profile.  These solutions correspond on the CFT side to an interface with a position-dependent  expectation value for a relevant operator and a source which jumps across the interface for a marginal operator.
 
 \end{abstract}

\setcounter{equation}{0}
\setcounter{footnote}{0}

\newpage

\section{Introduction}

Janus configurations   are solutions of supergravity theories which are dual to interface CFTs.  The original solution \cite{Bak:2003jk} was obtained by considering a  deformation of $\AdS_5\times S^5$ in type IIB supergravity where the dilaton has a nontrivial profile with respect to the slicing coordinate of an $\AdS_4$ slicing of $\AdS_5$. Subsequently, many more Janus solutions have been found in many different settings. One may distinguish two kinds of solutions: First, there are top-down constructions of Janus solutions in ten-dimensional type IIB or eleven-dimensional M-theory which preserve half of the supersymmetry. Such solutions are generically constructed by considering a warped product of $\AdS$ and sphere factors over a two-dimensional Riemann surface with boundary (see e.g.~\cite{DHoker:2007zhm,DHoker:2006qeo,DHoker:2008lup,DHoker:2009lky}). Second, there are solutions of gauged supergravities in lower dimensions with various amounts of broken and unbroken supersymmetries (see e.g.~\cite{Clark:2005te,Bobev:2013yra,Suh:2011xc,Chiodaroli:2011nr,Gutperle:2017nwo,Pilch:2015dwa,Karndumri:2016tpf,Bobev:2019jbi,Bobev:2020fon}). Solutions of the second kind are useful since holographic calculations of quantities such as the entanglement entropy, sources and expectation values of operators, and correlation functions in the Janus background are easier to perform in the lower-dimensional supergravity. In many cases, such solutions can be constructed as consistent truncations,  which can be lifted to solutions of   ten- or eleven-dimensional supergravity.

In the present paper, we consider a particular example of the second approach. We construct Janus solutions in three-dimensional $\cN=8$ gauged supergravity. Such theories are naturally related to $\AdS_3\times S^3\times M_4$ compactifications of type IIB, where $M_4$ is either  $T_4$ or $K3$.
We consider one of the simplest nontrivial settings where we find solutions which preserve eight of the sixteen supersymmetries of the $\AdS_3$ vacuum, where only two scalars in the coset have a nontrivial profile. One interesting feature of these solutions is that one scalar is dual to a marginal operator  with dimension $\Delta=2$ where the source terms  have different values on the two sides of the interface. This behavior  is the main feature of the original Janus solution \cite{Bak:2003jk,Clark:2004sb}. On the other hand,  the second scalar is 
dual to a relevant operator with dimension $\Delta=1$ with  a vanishing source term and  a position-dependent expectation value.  This behavior   is  a feature of the Janus solution in M-theory \cite{DHoker:2009lky}.

The structure of the paper is as follows: in section \ref{sec2} we review $\cN=8$ gauged supergravity in three dimensions, and in section \ref{sec3} we construct the half-BPS Janus solutions and investigate some of their properties using the AdS/CFT dictionary, including the calculation of the holographic entanglement entropy. We discuss some generalizations and directions for future research in section \ref{sec4}. Some technical details are relegated to appendix \ref{secappa}.

%%%%%%%%%%%%%%%%%%%%%%
\section{$d=3$, $\cN=8$ gauged supergravity}
\label{sec2}
%%%%%%%%%%%%%%%%%%%%%%

In the following, we will use the notation and conventions of \cite{Nicolai:2001ac}. 
The scalar fields of $d=3$, $\cN=8$ gauged supergravity are parameterized by  a $G/H = \SO(8, n)/\qty\big(\SO(8) \times \SO(n))$ coset, which has $8n$ independent scalar degrees of freedom.
This theory can be obtained by a truncation of six-dimensional $\cN = (2, 0)$ supergravity on $\AdS_3 \times S^3$ coupled to $n_T \geq 1 $ tensor multiplets, where $n_T = n-3$.
The cases $n_T = 5$ and $21$ correspond to compactifications of ten-dimensional type IIB on $T^3$ and $K3$, respectively. See \cite{Samtleben:2019zrh} for a discussion of  consistent truncations of six-dimensional $\cN = (1, 1)$ and $\cN = (2, 0)$ using exceptional field theory.   

For future reference, we use the following index conventions:
	\begin{itemize}
	\item $I, J, \dotsc = 1, 2, \dotsc, 8$ for $\SO(8)$.
	\item $r, s, \dotsc = 9, 10, \dotsc, n+8$ for $\SO(n)$.
	\item $\bar I, \bar J,\dotsc = 1, 2, \dotsc, n+8$ for $\SO(8, n)$.
	\item $\cM, \cN, \dotsc$ for generators of $\SO(8, n)$.
	\end{itemize}
	
Let the generators of $G$ be $\{t^\cM\} = \{t^{\bar I \bar J} \} = \{X^{IJ}, X^{rs}, Y^{Ir}\}$, where $Y^{Ir}$ are the non-compact generators.
Explicitly, the  generators of the vector representation are given by
	\begin{align} \tensor{(t^{\bar I \bar J})}{^{\bar K}_{\bar L}} = \eta^{\bar I \bar K} \delta^{\bar J}_{\bar L} - \eta^{\bar J \bar K} \delta^{\bar I}_{\bar L}
	\end{align}
where $\eta^{\bar I \bar J} = \diag( + + + + + + + + - \cdots)$ is the $\SO(8,n)$-invariant tensor.
These generators satisfy the following  commutation relations,
	\begin{align} [t^{\bar I \bar J}, t^{\bar K \bar L}] = 2\qty( \eta^{\bar I [\bar K} t^{ \bar L ] \bar J} - \eta^{\bar J [\bar K} t^{ \bar L ] \bar I} )
	\end{align}

The scalars fields can be parametrized by a $G$-valued matrix $L(x)$ in the vector representation, which transforms under $H$ and the gauge group $G_0\subseteq G$ by
	\begin{align} L(x) \longrightarrow g_0(x) L(x) h^{-1}(x)
	\end{align}
for $g_0 \in G_0$ and $h \in H$.
The Lagrangian is invariant under such transformations.
We can pick a $\SO(8) \times \SO(n)$ gauge to put the coset representative into symmetric gauge,
	\begin{align}  L = \exp( \phi_{Ir} Y^{Ir} ) \label{ldef}
	\end{align}
for scalar fields $\phi_{Ir}$.
The $\tensor{\cV}{^\cM_\cA}$ tensors are defined by
	\begin{align}  L^{-1} t^\cM L = \tensor{\cV}{^\cM_\cA} t^\cA = \frac{1}{2} \tensor{\cV}{^\cM_{IJ}} X^{IJ} + \frac{1}{2} \tensor{\cV}{^\cM_{rs}} X^{rs} + \tensor{\cV}{^\cM_{Ir}} Y^{Ir}
	\end{align}

The gauging of the supergravity is accomplished by introducing Chern-Simons gauge fields $B^\cM_\mu$ and  choosing an embedding tensor $\Theta_{\cM \cN}$ (which has to satisfy various identities \cite{deWit:2003ja}) that determines which isometries are gauged, the coupling to the Chern-Simons fields, and additional terms in the supersymmetry transformations and action depending on the gauge couplings. In the following, we will make one of the simplest choices and 
gauge a $G_0 = \SO(4)$ subset of  $\SO(8)$.
Explicitly, we further divide the $I, J$ indices into
	\begin{itemize}
	\item $i, j, \dotsc = 1, 2, 3, 4$ for $G_0 = \SO(4)$.
	\item $\bi, \bj, \dotsc = 5, 6, 7, 8$ for the remaining ungauged $\SO(4) \subset \SO(8)$.
	\end{itemize}
The embedding tensor  we will employ in the following  has the non-zero entries
	\begin{align} \label{embedt}
	\Theta_{IJ, KL} = \epsilon_{ijk\ell}
	\end{align}
As this is totally antisymmetric, the trace is $\theta = 0$. As discussed in \cite{Nicolai:2001ac}, this choice of embedding tensor produces a supersymmetric $\AdS_3$ ground state with 
\begin{align}
\SU(2|1,1)_L \times \SU(2|1,1)_R
\end{align}
super-algebra of isometries.
From the embedding tensor, the $G_0$-covariant currents can be obtained,
	\begin{align} L^{-1} (\partial_\mu + g \Theta_{\cM \cN} B_\mu^\cM t^\cN ) L = \frac{1}{2} \cQ^{IJ}_\mu X^{IJ} + \frac{1}{2} \cQ^{rs}_\mu X^{rs} + \cP^{Ir}_\mu Y^{Ir}
	\end{align}
It is convenient to define the $T$-tensor,
	\begin{align} T_{\cA | \cB} = \Theta_{\cM \cN} \tensor{\cV}{^\cM_\cA} \tensor{\cV}{^\cN_\cB}
	\end{align}
as well as the  tensors $A_{1, 2, 3}$ which will appear in the potential and the supersymmetry transformations.
	\begin{align}
	A_1^{AB} &= - \frac{1}{48} \Gamma^{IJKL}_{AB} T_{IJ|KL} \nonumber \\
	A_2^{A\dot A r} &= - \frac{1}{12} \Gamma^{IJK}_{A\dot A} T_{IJ|Kr} \nonumber \\
	A_3^{\dot A r \dot B s} &=  \frac{1}{48} \delta^{rs} \Gamma^{IJKL}_{\dot A \dot B} T_{IJ|KL} + \frac{1}{2} \Gamma^{IJ}_{\dot A \dot B} T_{IJ|rs}
	\end{align}
$A, B$ and $\dot A, \dot B$ are $\SO(8)$-spinor indices and our conventions for the $\SO(8)$ Gamma matrices are presented in appendix \ref{appa1}.

We take the spacetime signature $\eta^{ab} = \diag(+--)$ to be mostly negative.
The bosonic Lagrangian is
	\begin{align} 
	e^{-1} \cL &= - \frac{1}{4} R + \frac{1}{4} \cP_\mu^{Ir} \cP^{\mu\,Ir} + W - \frac{1}{4} e^{-1} \epsilon^{\mu\nu\rho} g \Theta_{\cM \cN} B_\mu^\cM \qty( \partial_\nu B_\rho^\cN + \frac{1}{3} g \Theta_{\cK \cL} \tensor{f}{^{\cN \cK}_{\cP}} B_\nu^\cL B_\rho^\cP ) \nonumber \\
	W &= \frac{1}{4} g^2 \qty( A^{AB}_1 A^{AB}_1 - \frac{1}{2} A^{A \dot A r}_2 A^{A \dot A r}_2 )  \label{lagrange}
	\end{align}
The SUSY transformations are
	\begin{align}
	\delta \chi^{\dot A r} &= \frac{1}{2} i \Gamma^I_{A\dot A} \gamma^\mu \epsilon^A \cP^{Ir}_\mu + g A^{A \dot A r}_2 \epsilon^A \nonumber \\
	\delta \psi^A_\mu &= \qty(\partial_\mu \epsilon^A + \frac{1}{4} \omega_\mu^{ab} \gamma_{ab} \epsilon^A + \frac{1}{4} \cQ^{IJ}_\mu \Gamma^{IJ}_{AB} \epsilon^B) + i g A^{AB}_1 \gamma_\mu \epsilon^B 
	\end{align}

\subsection{The $n=1$ case}

In this section we will  consider the  $n = 1$ theory, i.e.~the scalar fields lie in a $\SO(8,1)/\SO(8)$ coset. 
The reason for this is that the resulting expressions for the supersymmetry transformations and BPS conditions are compact and everything can be worked out in detail. 
Furthermore, we believe that this case illustrates the important features of more general solutions.

As the index $r=9$ takes only one value in this case, the scalar fields  in the coset representative (\ref{ldef})  are denoted by $\phi_i\equiv\phi_{i9}$ for $i=1,2,\dotsc,8$.  We define the following quantities for notational convenience,
	\begin{align}
	\Phi^2 &\equiv \phi_I \phi_I = \phi_1^2 + \phi_2^2 + \phi_3^2 + \phi_4^2 + \phi_5^2 + \phi_6^2 +  \phi_7^2 + \phi_8^2 \nonumber \\
	\phi^2 &\equiv \phi_i \phi_i = \phi_1^2 + \phi_2^2 + \phi_3^2 + \phi_4^2 \nonumber \\
	\bar{\phi}^2 &\equiv \phi_{\bi} \phi_{\bi} = \phi_5^2 + \phi_6^2 + \phi_7^2 + \phi_8^2 
	\end{align}
The components of the $\tensor{\cV}{^\cM_\cA}$ tensor are, with no summation over repeated indices and $I, J, K, L$ being unique indices,
	\begin{align}
	\tensor{\cV}{^{IJ}_{IJ}} &= 1 + (\phi_I^2 + \phi_J^2) \frac{\cosh \Phi - 1}{\Phi^2} & \tensor{\cV}{^{IJ}_{IK}} &= \phi_J \phi_K \frac{\cosh \Phi - 1}{\Phi^2} \nonumber \\
	\tensor{\cV}{^{IJ}_{KL}} &= 0 &  \tensor{\cV}{^{I9}_{I9}} &= \cosh \Phi - \phi_I^2 \frac{\cosh\Phi - 1}{\Phi^2} \nonumber \\
	\tensor{\cV}{^{I9}_{J9}} &= - \phi_I \phi_J \frac{\cosh\Phi - 1}{\Phi^2} & \tensor{\cV}{^{IJ}_{I9}} &= \tensor{\cV}{^{I9}_{IJ}} = \phi_J \frac{\sinh \Phi}{\Phi} \nonumber \\
	\tensor{\cV}{^{IJ}_{K9}} &= \tensor{\cV}{^{K9}_{IJ}} = 0 
	\end{align}
The $u$-components of the $\cQ^{IJ}_\mu$ and $\cP^{I}_\mu$ tensors are
	\begin{align} 
	\cQ_u^{IJ} &= (\phi_I' \phi_J - \phi_I \phi_J') \frac{\cosh \Phi - 1}{\Phi^2} + g \Theta_{\cM \cN} B^{\cM}_u \cV^{\cN}_{IJ} \nonumber \\
	\cP_u^{I} &= \phi_I' \frac{\sinh \Phi}{\Phi} - \phi_I \Phi' \frac{\sinh \Phi - \Phi}{\Phi^2} + g \Theta_{\cM \cN} B^{\cM}_u \cV^{\cN}_{I9} \label{qandp}
	\end{align}
where the prime $' \equiv \pdv*{u}$ denotes the derivative with respect to $u$.
The terms involving the gauge field have different forms depending on whether $I, J$ are in $i$ or $\bi$.
	\begin{align}
	\Theta_{\cM \cN} B^{\cM}_u \cV^{\cN}_{ij} &= \epsilon_{ijk\ell} \qty[ \frac{1}{2} B^{k\ell}_u \qty( 1 + (\phi_i^2 + \phi_j^2) \frac{\cosh \Phi - 1}{\Phi^2} ) + \qty(\phi_i B^{ik}_u\phi_\ell + \phi_j B^{jk}_u\phi_\ell) \frac{\cosh \Phi - 1}{\Phi^2} ] \nonumber \\
	\Theta_{\cM \cN} B^{\cM}_u\cV^{\cN}_{i\bi} &=  \frac{1}{2} \epsilon_{ijk\ell} \phi_{\bi} \phi_j B^{k\ell}_u\frac{\cosh \Phi - 1}{\Phi^2} \nonumber \\
	\Theta_{\cM \cN} B^{\cM}_u \cV^{\cN}_{\bi\bj} &=  0 \nonumber \\
	\Theta_{\cM \cN} B^{\cM}_u \cV^{\cN}_{i9} &=  \frac{1}{2} \epsilon_{ijk\ell} \phi_j B^{k\ell}_u \frac{\sinh \Phi}{\Phi} \nonumber \\
	\Theta_{\cM \cN} B^{\cM}_u \cV^{\cN}_{\bi9} &=  0 
	\end{align}
	The $T$-tensor has non-zero components
	\begin{align}
	T_{ij|k\ell} &= \epsilon_{ijk\ell} \qty(\phi^2 \frac{\cosh \Phi -1 }{\Phi^2} + 1) \nonumber \\
	T_{ij|k \bi} &= \epsilon_{ijk\ell} \phi_\ell \phi_{\bi} \frac{\cosh \Phi - 1}{\Phi^2} \nonumber \\
	T_{ij|k9} &= \epsilon_{ijk\ell} \phi_\ell \frac{\sinh \Phi}{\Phi}
	\end{align}
Taking $\epsilon_{1234} = 1$, we can use the $T$-tensor to compute
	\begin{align} 
	A_1^{AB} &= - \frac{1}{2} \Gamma^{1234}_{AC} \Bigg[ \qty(\phi^2 \frac{\cosh \Phi -1 }{\Phi^2} + 1) \delta_{CB} + (\Gamma^i_{C \dot A} \phi_i) (\Gamma^{\bi}_{\dot A B} \phi_{\bi}) \frac{\cosh \Phi - 1}{\Phi^2} \Bigg] \nonumber \\
	A_2^{A \dot A} &= - \frac{1}{2} \Gamma^{1234}_{A B} (\Gamma^i_{B \dot A} \phi_i) \frac{\sinh \Phi}{\Phi} \nonumber \\
	A_3^{\dot A \dot B} &= - A_1^{AB} \delta_{A \dot A} \delta_{B \dot B}
	\end{align}
Note that $A_1^{AB} = A_1^{BA}$ and 
	\begin{align}
	A_1^{AC} A_1^{BC} &= \frac{1}{4} \delta_{AB}  \qty(  \frac{\phi^2\sinh^2\Phi}{\Phi^2} + 1 ) \nonumber \\
	A_2^{A\dot A} A_2^{B \dot A} &= \frac{1}{4} \delta_{AB} \frac{\phi^2 \sinh^2\Phi}{\Phi^2}  \label{eq:Asq}
	\end{align}
so the scalar potential (\ref{lagrange}) becomes
	\begin{align}
	W = \frac{g^2}{4}  \qty(  \frac{\phi^2\sinh^2 \Phi}{\Phi^2} + 2 )
	\end{align}

\section{Half-BPS Janus solutions}
\label{sec3}

In this section, we construct Janus solutions which preserve eight  of the sixteen supersymmetries of the  $\AdS_3$ vacuum.  Our strategy is to use an $\AdS_2$ slicing of $\AdS_3$ and make the scalar fields as well as the metric functions only dependent on the slicing coordinate.  One complication is given by the presence of the gauge fields; due to the Chern-Simons action, the only consistent Janus solution will have vanishing field  strength. We show that the gauge fields can be consistently set to zero for our solutions.

\subsection{Janus ansatz}
	
We take the Janus ansatz for the metric, scalar fields and Chern-Simons gauge fields,
	\begin{align} 
	\dd{s^2} &= e^{2B(u)} \qty(\frac{ \dd{t^2} - \dd{z^2} }{z^2}) - e^{2D(u)} \dd{u}^2 \nonumber \\
	\phi_I &= \phi_I(u) \nonumber \\
	B^\cM &= B^\cM(u) \dd{u}
	\end{align}
The $\AdS_3$ vacuum solution given by $\phi_I \equiv 0$  and $e^{B}= e^{D} = L \sec u$ has a curvature radius  related to the coupling constant by $L^{-1} = g$.
The spin connection 1-forms are
	\begin{align}
	\omega^{01} &= \frac{\dd{t}}{z} & \omega^{02} &= -\frac{B'e^{B-D}}{z} \dd{t} & \omega^{12} &= -\frac{B'e^{B-D}}{z} \dd{z} 
	\end{align}
so the gravitino supersymmetry  variation  $\delta \psi^A_\mu = 0$ is
	\begin{align}
	0 &= \partial_t \epsilon + \frac{1}{2z} \gamma_0 \qty(\gamma_1  - B' e^{B-D} \gamma_2 + 2 i g e^B A_1) \epsilon \nonumber \\
	0 &= \partial_z \epsilon + \frac{1}{2z} \gamma_1 \qty( - B' e^{B-D} \gamma_2 + 2 i g e^B A_1 ) \epsilon \nonumber \\
	0 &= \partial_u \epsilon + \frac{1}{4} \cQ_u^{IJ} \Gamma^{IJ} \epsilon + i g e^D \gamma_2 A_1 \epsilon 
	\end{align}
where we have suppressed the $\SO(8)$-spinor indices.
As shown in appendix \ref{appa2}, the integrability conditions are
	\begin{align}
	0 &= \qty(1 -(2 g e^B A_1)^2 + (B' e^{B-D})^2 )\epsilon \nonumber \\
	0 &= 2ig e^B \qty( A_1' - \frac{1}{4} [ A_1, \cQ_u^{IJ} \Gamma^{IJ}] ) \epsilon + \qty( - \dv{u} (B' e^{B-D}) + (2 g e^B A_1)^2 e^{D-B} ) \gamma_2 \epsilon 
	\end{align}
The first integrability condition gives a first-order equation which must be true for all $\epsilon$, using the replacement for $A_1^2$ in (\ref{eq:Asq}),
	\begin{align}
	  0 = 1 - g^2 e^{2B} \qty(  \frac{\phi^2\sinh^2\Phi}{\Phi^2} + 1) + (B' e^{B-D})^2   \label{eq:box1}
	\end{align}
The derivative of this simplifies the second integrability condition to
	\begin{align}
	0 = \qty( A_1' - \frac{1}{4} [ A_1, \cQ_u^{IJ} \Gamma^{IJ}] ) \epsilon + \frac{i g e^D}{4 B'} \dv{u} \qty( \frac{\phi^2 \sinh^2 \Phi}{\Phi^2}) \gamma_2 \epsilon \label{eq:integ}
	\end{align}
The BPS equation $\smash{\delta \chi^{\dot A} = 0}$ is
	\begin{align}
	\qty(- \frac{i}{2} e^{-D} \Gamma^I \cP_u^{I} \gamma_2 + g A_2 )_{A\dot A} \epsilon^A = 0 
	\end{align}
When $g A_2^2 \neq 0$, this equation can be rearranged into the form of a projector
	\begin{align}\label{eq:proj}
	0 &= \qty( i M_{AB} \gamma_2 + \delta_{AB} ) \epsilon^A 
	\end{align}
	where $M_{AB}$ is given by
	\begin{align}
	  M_{AB} &= \frac{e^{-D}}{g} \frac{\Phi}{\phi^2 \sinh \Phi} (\Gamma^I_{A\dot A} \cP_u^I) (\Gamma^i_{\dot A C} \phi_i)  \Gamma^{1234}_{CB}
	\end{align}
For consistency of the projector, we must have
	\begin{align}
	M_{AB} M_{BC} = \delta_{AC}
	\end{align}
As $M^2 = 1$, every generalized eigenvector of rank $\geq 2$ is automatically an eigenvector, so $M$ is diagonalizable and has eight eigenvectors with eigenvalues $\pm 1$.
$M$ is traceless as it is a sum of products of 2 or 4 Gamma matrices, so it has an equal number of $+1$ and $-1$ eigenvectors.
The  operator  $ i M_{AB} \gamma_2$ in  the projector (\ref{eq:proj})  squares to one and is traceless, and projects onto an eight-dimensional space of unbroken supersymmetry generators. If this is the only projection imposed on the solution, it will be half-BPS and hence preserve eight of the sixteen supersymmetries of the vacuum.

The condition $M^2 = 1$ gives an equation first-order in derivatives of scalars.
	\begin{align}
	M^2 = \qty( \frac{e^{-D}\Phi}{g \phi^2 \sinh\Phi} )^2 \Big( & \phi^2 (-\cP_u^i \cP_u^i + \cP_u^{\bi}\cP_u^{\bi}) - 2 \phi^2 (\Gamma^{\bi}\cP_u^{\bi})(\Gamma^i \cP_u^i) \nonumber \\
	& + 2 (\cP_u^j \phi_j) (\Gamma^{\bi}\cP_u^{\bi} + \Gamma^{i}\cP_u^{i})(\Gamma^k \phi_k)\big)
	\end{align}
For this to be proportional to the identity, we need all $\Gamma^{\bi} \Gamma^i$ and $\Gamma^i \Gamma^j$ terms to vanish.
Vanishing of the latter requires us to impose the condition
	\begin{align}
	 \cP^i_u \phi_j = \cP^j_u \phi_i   \label{eq:box2}
	\end{align}
As the ratio $\cP_u^i/\phi_i$ is the same for all $i$, this implies 
	\begin{align}
	\sum_i \cP_u^i \phi_i = \sum_i \frac{\cP_u^i}{\phi_i} \phi_i^2 = \frac{\cP_u^1}{\phi_1} \phi^2 \qquad\implies\qquad - \phi^2 \cP_u^i + \phi_i \sum_j \cP_u^j \phi_j = 0
	\end{align}
This means that imposing Eq.~(\ref{eq:box2}) also ensures that the $\Gamma^{\bi} \Gamma^i$ terms vanish.
Note that
	\begin{align}
	\sum_i \cP_u^i \cP_u^i = \sum_i \frac{\cP_u^i}{\phi_i} \frac{\cP_u^i}{\phi_i} \phi_i^2 = \qty(\frac{\cP_u^1}{\phi_1})^2 \phi^2 
	\end{align}
so the $M^2 = 1$ condition becomes
	\begin{align}
	M^2 =   \qty( \frac{e^{-D}\Phi}{g \phi^2 \sinh\Phi} )^2 \phi^2 (\cP_u^i \cP_u^i + \cP_u^{\bi}\cP_u^{\bi}) = 1 \label{eq:box3}
	\end{align}

We now give the argument why the Chern-Simons gauge fields  can be set to zero. 
Since we demand that the $B^{\cM}_\mu$ only has a component along the $u$ direction and only depends on $u$, the field strength vanishes, consistent with the equation of motion coming from the variation of the Chern-Simons term in the action (\ref{lagrange}) with respect to the gauge field. However, there is another term which contains the gauge field, namely the kinetic term of the scalars via (\ref{qandp}).  For the gauge field to be consistently set to zero, we have to impose
	\begin{align}
	\left. {\delta \cL \over \delta B^{k\ell}_u} \right|_{B^{\cM}_u=0} =0
	\end{align}	
For the Janus ansatz, we find
	\begin{align}
	\left. {\delta \cL \over \delta B^{k\ell}_u} \right|_{B^{\cM}_u=0} = e g \epsilon_{ijk\ell} \cP^{i\, u} \phi_j {\sinh \Phi \over \Phi}  
 	\end{align}
which indeed vanishes due to Eq.~(\ref{eq:box2}) imposed by the half-BPS condition. 

For a half-BPS solution, the second integrability condition (\ref{eq:integ}) should be identical to the projector (\ref{eq:proj}).
Indeed, we have the simplification
	\begin{align}
	A_1' - \frac{1}{4} & [ A_1, \cQ_u^{IJ} \Gamma^{IJ} ] = - \frac{1}{2} \frac{\phi^2 \sinh^2 \Phi}{\Phi^2} M^\top
	\end{align}
so the Gamma matrix structures of the two equations match.
Equating the remaining scalar magnitude gives us an equation for the metric factor $e^B$,
	\begin{align}
	-B' = \dv{u} \ln \frac{\phi \sinh \Phi}{\Phi} \label{eq:box4}
	\end{align}

We can now solve for the metric.
Let us define
	\begin{align}
	\alpha(u) \equiv \frac{\phi \sinh \Phi}{\Phi}
	\end{align}
and set the integration constant for $B$ to be
	\begin{align} 
	e^{B} = \frac{|C|}{g\alpha}
	\end{align}
Plugging this into the first integrability condition (\ref{eq:box1}) and picking the gauge $e^{-D} \equiv g$, we have a first-order equation for $\alpha$,
	\begin{align}
	0 = \alpha^2 - C^2 (\alpha^2 + 1 - \alpha'^2 / \alpha^2)
	\end{align}
The solution depends on the value of $C \in [0, 1]$ and up to translations in $u$ is
	\begin{align}
	\alpha &= e^{\pm u} & & \text{if } C = 1\nonumber \\
	\alpha &= \frac{|C|}{\sqrt{1 - C^2}} \sech u & & \text{if } 0 \leq  C < 1 \label{eq:alpha}
	\end{align}
We will take the case $0 \leq C < 1$.
This implies that the metric is
	\begin{align}\label{metjan}
	\dd{s^2} = g^{-2} \qty[ (1 - C^2) \cosh^2 u \qty( \frac{\dd{t^2} - \dd{z^2}}{z^2} )  - \dd{u^2} ]
	\end{align}
The choice $C = 0$ corresponds to the $\AdS_3$ vacuum.

\subsection{$\phi_4, \phi_5$ truncation}\label{sec32}

We have yet to fully solve the half-BPS conditions (\ref{eq:box2}) and (\ref{eq:box3}).
For simplicity, let us consider the case where only $\phi_4, \phi_5$ are non-zero and the other scalars are identically zero, which trivially satisfies Eq.~(\ref{eq:box2}).
It turns out that the important features of the Janus solution are captured by this truncation.

We introduce the following abbreviations
	\begin{align}
	\Phi^2 &= \phi_4^2 + \phi_5^2 & \phi &= |\phi_4| & \bar{\phi} &= |\phi_5|
	\end{align}
Let us define
	\begin{align}
	\beta(u) &\equiv \frac{\phi_5 \sinh \Phi}{\Phi}
	\end{align}
so that
	\begin{align}
	\alpha^2 + \beta^2 &= \sinh^2\Phi \nonumber \\
	\cP^4_u &= \alpha' + \alpha \Phi' \frac{1 - \cosh\Phi}{\sinh \Phi} \nonumber \\
	\cP^5_u &= \beta' + \beta \Phi' \frac{1 - \cosh\Phi}{\sinh \Phi}
	\end{align}
Plugging these into Eq.~(\ref{eq:box3}) simplifies to
	\begin{align}
	\alpha'^2 + \beta'^2 - \frac{(\alpha' \alpha + \beta' \beta)^2}{1 + \alpha^2 + \beta^2} &= \alpha^2
	 	\end{align}
This can be rearranged into a first-order equation in $f \equiv \beta / \sqrt{1+\alpha^2}$,
	\begin{align}
	f' = \frac{\alpha^2/C}{1 + \alpha^2} \sqrt{1 + f^2}
	\end{align}
where a sign ambiguity from taking a square-root has been absorbed into $C$, which is now extended to $C \in (-1, 1)$.
Using the explicit solution (\ref{eq:alpha}) for $\alpha$, by noting that 
	\begin{align}
	\dv{u} \tanh^{-1}( C \tanh u) = \frac{C \sech^2 u}{1 - C^2 \tanh^2 u} = \frac{\alpha^2/C}{1+\alpha^2}
	\end{align}
the general solution is
	\begin{align}
	f(u) &= \frac{\sinh p + C \cosh p \tanh u}{\sqrt{1 - C^2 \tanh^2 u}} \nonumber \\
	\beta(u) &= \frac{1}{\sqrt{1 - C^2}} (\sinh p + C \cosh p \tanh u)  \label{eq:beta}
	\end{align}
for some constant $p \in \mathbb{R}$.
For later convenience, we also redefine $C = \tanh q$ for $q \in \mathbb{R}$.

In summary, we have solved for the scalars $\phi_4, \phi_5$ implicitly through the functions $\alpha, \beta$,
	\begin{align} 
	\frac{|\phi_4| \sinh \Phi}{\Phi} &= |\sinh q| \sech u \nonumber \\
	\frac{\phi_5 \sinh \Phi}{\Phi} &=  \sinh p \cosh q + \cosh p \sinh q \tanh u \label{eq:scalar-sol}
	\end{align}
for real constants $p, q$.
Note that the reflection $\phi_4 \to -\phi_4$ also gives a valid solution.
We have explicitly checked that the Einstein equation and scalar equations of motion are satisfied.

The $\phi_4$ scalar goes to zero at $u = \pm \infty$ as it is a massive scalar degree of freedom, and has a sech-like profile near the defect. 
The $\phi_5$ scalar interpolates between two boundary values at $u = \pm \infty$, and has a tanh-like profile.
The constant $p$ is related to the boundary values of the $\phi_5$ scalar, as we can note that
	\begin{align}
	\phi_5(\pm \infty) =  p \pm q
	\end{align}
The constant $q$ is then related to the jump value of the $\phi_5$ scalar.
The defect location $u = 0$ can also be freely translated to any point along the axis.
Below is a plot of the solution for the choice $(p, q) = (0, 1)$.

\begin{figure}[htbp]
\begin{center}
{\centerline{\includegraphics[width=9cm]{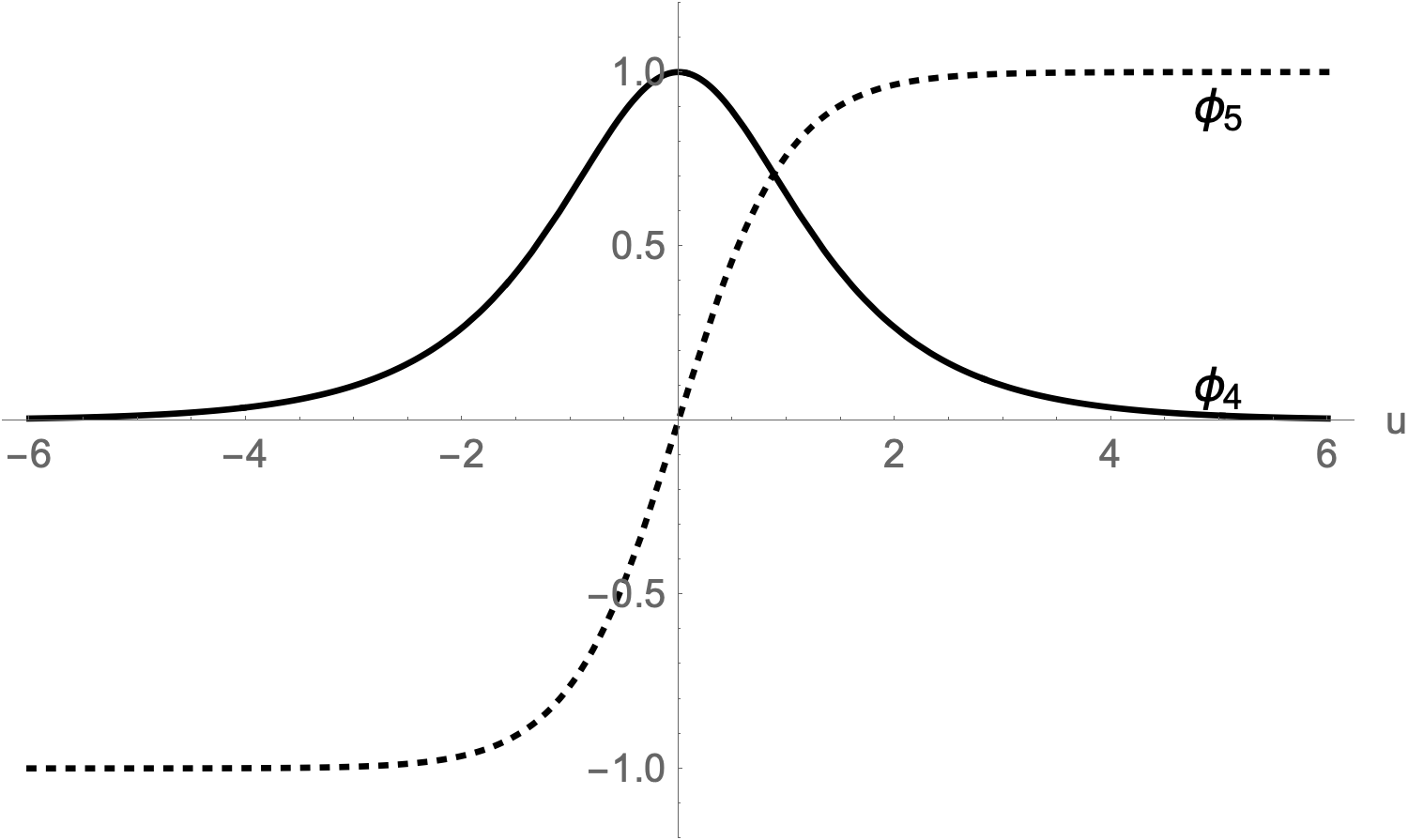}}}

\caption{Plot of $\phi_4$ and $\phi_5$ for  $(p, q) = (0, 1)$ }
\label{default}
\end{center}
\end{figure}

\subsection{Holography}

In our AdS-sliced coordinates, the boundary is given by the two $\AdS_2$ components at $u = \pm \infty$, which are joined together at the $z=0$ interface.
Using $C = \tanh q$, the metric (\ref{metjan}) becomes
	\begin{align} 
	\dd{s^2} = g^{-2} \qty[ \sech^2 q \cosh^2 u \qty( \frac{\dd{t^2} - \dd{z^2}}{z^2} )  - \dd{u^2} ] \label{eq:metric}
	\end{align}
Note that this is not $\AdS_3$ unless  $q=0$, which corresponds to the vacuum solution with all scalars vanishing. The  spacetime is, however, asymptotically $\AdS_3$. In the limit of $u\to \pm \infty$, the $\sech^2 q$ can be eliminated from the leading $e^{\pm 2 u}$ term in the metric  (\ref{eq:metric}) by a coordinate shift. We will present the asymptotic mapping to a Fefferman-Graham (FG) coordinate system below. In the following, we will set the $\AdS$ length scale to unity for notational simplicity, i.e.~$g \equiv 1$.

According to the AdS/CFT correspondence, the mass $m^2$ of a supergravity scalar field in $d=3$ is related to the scaling dimension $\Delta$ of the dual CFT operator by
	\begin{align} m^2 = \Delta (\Delta - 2) 
	\end{align}
This relation comes from the linearized equations of motion for the scalar field near the asymptotic $\AdS_3$ boundary.
Expanding the supergravity action (\ref{lagrange}) to quadratic order around the $\AdS_3$ vacuum shows that the $\phi_4$ field has mass $m^2 = -1$, so the dual operator is relevant with $\Delta = 1$ and saturates the Breitenlohner-Freedman (BF) bound \cite{Breitenlohner:1982bm}.  Note that we choose the standard quantization \cite{Klebanov:1999tb}, which is the correct one for a supersymmetric solution.
The $\phi_5$ field is massless, so the dual CFT operator is marginal with  scaling dimension $\Delta = 2$.

In FG coordinates,\footnote{The $\AdS_3$ metric in Poincar\'{e} coordinates is
\[ \dd{s^2} = \frac{-\dd{\rho^2} + \dd{t^2} - \dd{x^2}}{\rho^2} \]
and is related to the AdS-sliced metric by the coordinate change
\begin{align*}
z &= \sqrt{x^2 + \rho^2} & \sinh u &= x / \rho
\end{align*}
}  the general expansion for a scalar field near the asymptotic $\AdS_3$ boundary at $\rho = 0$ is
	\begin{align} 
	\phi_{\Delta=1}&\sim \psi_0  \, \rho \ln \rho + \phi_0  \, \rho + \cdots  \nonumber \\
	\phi_{\Delta\neq 1} &\sim \tilde \phi_{0} \, \rho^{2 - \Delta} + \tilde \phi_{2} \, \rho^\Delta  +\cdots
		\end{align}
Since $\phi_{\Delta=1}$ saturates the BF bound, holographic renormalization and the holographic dictionary are subtle due to the presence of the logarithm \cite{Witten:2001ua}.  As we show  below  for the solution (\ref{eq:scalar-sol}), there is no logarithmic term present and $\phi_0$ can be identified with the expectation value of the dual operator \cite{Witten:2001ua,Marolf:2006nd}.  For the massless field $\phi_{\Delta=2}$, we can identify $\tilde{\phi}_{0}$ with the source and $ \tilde{\phi}_{2}$ with the expectation value of the dual operator.

It is difficult to find a global  map which puts the metric (\ref{eq:metric}) in FG form.  Here, we limit our discussion to the coordinate region away from the defect, where we take  $u \to \pm \infty$ and keep $z$ finite  \cite{Papadimitriou:2004rz,Jensen:2013lxa}. This limit  probes the region away from the interface on the boundary. The coordinate change suitable for the $u \to \infty$ limit can be expressed as a power series,
	\begin{align}
	z &= x + \frac{\rho^2}{2 x} + \cO(\rho^4) \nonumber \\
	e^u &= \cosh q \qty( \frac{2 x}{\rho} + \frac{\rho}{2 x} + \cO(\rho^3) ) \label{eq:asymp-coordchange}
	\end{align}
The metric becomes
	\begin{align}
	\dd{s^2} &= \frac{1}{\rho^2} \qty[ -\dd{\rho}^2 + \qty(1 - \frac{\rho^2\tanh^2 q}{2 x^2})(\dd{t}^2 -\dd{x^2}) + \cO(\rho^3)]
	\end{align}
In the $u \to -\infty$ limit, the asymptotic form of the metric is the same and the coordinate change is (\ref{eq:asymp-coordchange}) with the replacements $e^u \to e^{-u}$ and $x \to -x$.

Using this coordinate change, the expansions of the scalar fields near the boundary is
\begin{align}
	|\phi_4| &= |\tanh q| \frac{p + \tilde q}{\sinh(p + \tilde q)} \cdot \frac{\rho}{|x|} + \cO(\rho^3) \nonumber \\
	\phi_5 &= (p + \tilde q) - \frac{1}{2\sinh(p + \tilde q)} \qty(\frac{p + \tilde q}{\sinh(p + \tilde q)} \tanh^2 q  + \frac{\sinh p \tanh {\tilde q}}{\cosh q} ) \cdot \frac{\rho^2}{x^2} + \cO(\rho^4)
	\end{align}
where $\tilde{q} \equiv q x / |x|$ (see appendix \ref{appa3} for details).
The defect is located on the boundary at $x = 0$.
We can see that the relevant operator corresponding to $\phi_4$ has no term proportional to $\rho \ln \rho$ in the expansion. This implies that the source is zero and the dual operator has a position-dependent expectation value.
The marginal operator corresponding to $\phi_5$ has a source term which takes different values on the two sides of the defect, corresponding to a Janus interface where the modulus associated with the marginal operator jumps across the interface.

Another  quantity  which can be calculated  holographically is the entanglement entropy for an interval $A$ using the Ryu-Takanayagi prescription \cite{Ryu:2006bv},
\begin{align}
S_{\rm EE}={ {\rm Length } (\Gamma_A) \over 4 G_N^{(3)}}
\end{align}
where $\Gamma_A$ is the minimal curve  in the bulk which ends on $\partial A$.

\begin{figure}[htbp]
\begin{center}
{\centerline{\includegraphics[width=13cm]{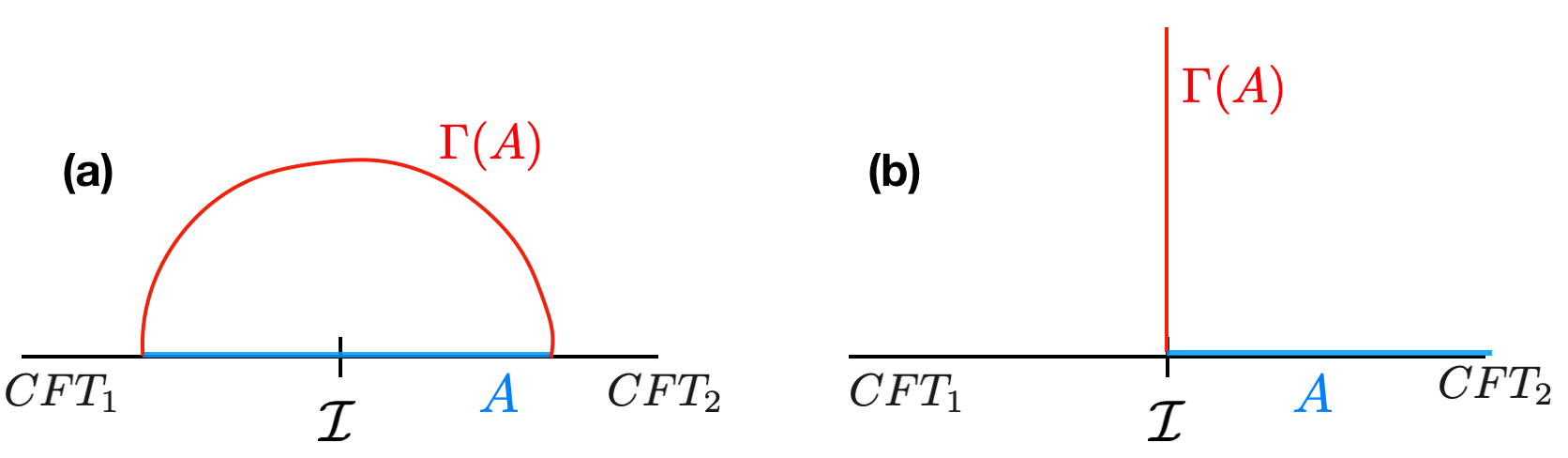}}}
\caption{(a) The entagling surface $A$ is symmetric around the interface ${\cal I}$, (b)  The entagleing surface $A$ is ends at the interface ${\cal I}$}
\label{fig:fig2}
\end{center}
\end{figure}

 There are two qualitatively different choices for location of the interval in an interface CFT, as shown in figure \ref{fig:fig2}. First,   the interval can be chosen symmetrically around the defect \cite{Azeyanagi:2007qj,Chiodaroli:2010ur}. The minimal surface for such a symmetric interval is  particularly simple in the  $\AdS$-sliced coordinates (\ref{eq:metric}),  and is given by $z=z_0$ and $u\in (-\infty, \infty)$. The regularized length is given by
 \begin{align}
  {\rm Length  } (\Gamma_A)  = \int \dd{u} = u_{\infty}-u_{-\infty}
 \end{align} 
We can use (\ref{eq:asymp-coordchange}) to relate the FG cutoff $\rho=\epsilon$, which furnishes the UV cutoff on the CFT side,  to the cutoff $u_{\pm \infty}$ in the $\AdS$-sliced metric,
 \begin{align}
 u_{\pm \infty} = \pm \qty\big(- \log \epsilon + \log (2 z_0) +\log(\cosh q) )
 \end{align}
 Putting this together and using the expression for the central charge in terms of $G_N^{(3)}$ gives
 \begin{align}
 S_{\rm EE} = {c\over 3} \log {2z_0\over \epsilon} + {c\over 3} \log (\cosh q)
 \end{align}
 
Note that the first logarithmically divergent term is the standard expression for the entanglement entropy for a CFT without an interface present \cite{Calabrese:2004eu}, since $2 z_0$ is the length of the interval. The constant term  is universal in the presence of an interface  and can be interpreted as the defect entropy (sometimes called g-factor \cite{Affleck:1991tk}) associated with the interface.

Second, we can consider an interval which lies on one side of the interface and borders the interface \cite{Sakai:2008tt,Brehm:2015lja}.  	As shown in \cite{Gutperle:2015hcv}, the entangling surface is located at $u=0$ and the entanglement entropy  for an interval of length $l$ bordering the  interface is given by
\begin{align}
S'_{\rm EE} = {c\over 6}  \sech{q} \log {l \over \epsilon}
\end{align} 
	
\subsection{All scalars}

For completeness, we also present the general solution with all $\phi_I$ scalars turned on. 
Let us define
	\begin{align}
	\alpha_i(u) &\equiv \frac{\phi_i \sinh \Phi}{\Phi} & i &= 1, 2, 3, 4 \nonumber \\
	\beta_{\bi}(u) &\equiv \frac{\phi_{\bi} \sinh\Phi}{\Phi} & \bi &= 5, 6, 7, 8
	\end{align}
As a consequence of Eq.~(\ref{eq:box2}), the ratio $\phi_i'/\phi_i$ is the same for all $i$ so all the $\phi_i$ scalars are proportional to each other.
In other words, we have $\alpha_i = n_i \alpha$ for constants $n_i$ satisfying $n_i n_i = 1$, where $\alpha$ is given in Eq.~(\ref{eq:alpha}).
Then Eq.~(\ref{eq:box3}) becomes
	\begin{align} 
	\alpha'^2 + \beta_{\bi}' \beta_{\bi}' - \frac{(\alpha' \alpha + \beta_{\bi}' \beta_{\bi})^2}{1 + \alpha^2 + \beta_{\bi} \beta_{\bi} } = \alpha^2  \label{eq:box3-all}
	\end{align}
We can note that there exists a family of solutions where all $\beta_{\bi}$ functions satisfy
	\begin{align}
	\beta_{\bi} = n_{\bi} \beta
	\end{align}
for some function $\beta$ and constants $n_{\bi}$ satisfying $n_{\bi} n_{\bi} = 1$.
When this is the case, Eq.~(\ref{eq:box3-all}) then further simplifies to
	\begin{align}
	\alpha'^2 + \beta'^2 - \frac{(\alpha' \alpha + \beta' \beta)^2}{1 + \alpha^2 + \beta^2 } = \alpha^2
	\end{align}
which has already been solved in the previous section.
We can prove that these are the only solutions to Eq.~(\ref{eq:box3-all}) which satisfy the equations of motion.
The scalar dependence of the Lagrangian is
	\begin{align} e^{-1} \cL &\supset -\frac{g^2}{4} \cP^I_u \cP^I_u + W \nonumber \\
	&= -\frac{g^2}{4} \qty(\alpha'^2 + \beta_{\bi}' \beta_{\bi}' - \frac{(\alpha' \alpha + \beta_{\bi}' \beta_{\bi})^2}{1 + \alpha^2 + \beta_{\bi} \beta_{\bi} } - (\alpha^2 + 2)	 ) 
	\intertext{If we write the $\beta_{\bi}$ in spherical coordinates, where we call the radius $\beta$, this becomes}
	&= -\frac{g^2}{4} \qty(\alpha'^2 + \beta'^2 + \beta^2 K^2 - \frac{(\alpha' \alpha + \beta' \beta)^2}{1 + \alpha^2 + \beta^2 } - (\alpha^2 + 2)	 ) 
	\end{align}
where $K^2$ is the kinetic energy of the angular coordinates.\footnote{Explicitly, let $K^2 = \theta'^2 + \sin^2\theta \, \phi'^2 + \sin^2\theta \sin^2\phi \, \psi'^2$.}
We can treat $\alpha, \beta$, and the three angles as the coordinates of this Lagrangian.
The equation of motion from varying the Lagrangian with respect to $\alpha$ will only involve $\alpha$ and $\beta$ and their derivatives.
Plugging-in (\ref{eq:alpha}) for $\alpha$, satisfying this equation of motion fixes the form of $\beta$ to be what was found previously in Eq.~(\ref{eq:beta}).
This means that Eq.~(\ref{eq:box3-all}) simplifies to $\beta^2 K^2 = 0$ and the three angles must be constant. 

Therefore, the general solution is
	\begin{align}
	\frac{\phi \sinh \Phi}{\Phi} &= |\sinh q| \sech u \nonumber \\
	\beta &=  \sinh p \cosh q + \cosh p \sinh q \tanh u \nonumber \\
	\phi_i &= n_i \phi \qquad,\quad n_i n_i = 1 \nonumber \\
	\frac{\phi_{\bi} \sinh \Phi}{\Phi} &= n_{\bi} \beta  \qquad,\quad n_{\bi} n_{\bi} = 1
	\end{align}

\section{Discussion}
\label{sec4}

In this paper, we have presented Janus solutions for  $d=3$, ${\cal N}=8$ gauged supergravity. We constructed the simplest solutions with the smallest number of scalars, namely the $\SO(n,1)/\SO(8)$ coset. The solutions we found have  only two scalars displaying a nontrivial profile.  One scalar is dual to a marginal operator $O_2$ with scaling dimension $\Delta=2$ and the other scalar is dual to a relevant operator  $O_1$ with scaling dimension $\Delta=1$. We used  the holographic correspondence to  find the dual CFT  interpretation of these solutions. It is given by a superconformal  interface, with a constant source of the operator $O_2$ which jumps across the interface. For the operator  $O_1$, the source vanishes but there is an expectation value which depends on the distance from the interface. It would be interesting to study whether half-BPS Janus interfaces which  display these characteristics can be constructed in the two-dimensional $\cN=(4,4)$ SCFTs.

 We considered   solutions for the $\SO(n,1)/\SO(8)$ coset, but these solutions  can be trivially embedded into  the $\SO(8, n)/\qty\big(\SO(8) \times \SO(n))$ cosets with $n>1$. Constructing solutions with more scalars with nontrivial profiles  is in principle possible, but the explicit expressions for the quantities involved in the BPS equations are becoming very complicated. 
 We also believe that the $n=1$ case already illustrates the important features of the more general $n>1$ cosets. Another possible generalization is given  by considering more general gaugings. One important example is given by replacing the embedding tensor (\ref{embedt}) with
\begin{align}
 \Theta_{IJ, KL} = \epsilon_{ijk\ell}+ \alpha \epsilon_{\bi \bj \bar k \bar \ell}
	\end{align}
This is a deformation produces  an $\AdS_3$ vacuum which is dual to a SCFT  with a  large $D^1(2,1;\alpha)\times D^1(2,1;\alpha)$ superconformal algebra.   As discussed in \cite{Nicolai:2001ac}, this gauging is believed to be a truncation type II supergravity compactified on
 $\AdS_3\times S^3\times S^3\times S^1$  \cite{deBoer:1999gea,Gukov:2004ym}. It should be straightforward to adapt the methods  for finding solutions developed in the present paper to this case. 
 
 We calculated the holographic defect entropy  for our solution. It would be interesting to investigate whether this quantity can be related to the Calabi diastasis function following \cite{Bachas:2013nxa,DHoker:2014qtw}. For this identification to work we would have to consider the case $n=2$ for which the scalar coset is a K\"ahler manifold. 
 
 We leave these interesting questions for future work.

\section*{Acknowledgements}

We would like to thank Matteo Vicino for collaboration at the initial stages of this work and Per Kraus for useful conversations.
The work of M.~G.~was supported, in part, by the National Science Foundation under grant PHY-19-14412. 
K.~C.~and   M.~G.~are grateful to the Mani L.~Bhaumik Institute for Theoretical Physics for support.

\newpage

 \appendix
 
 \section{Technical details} \label{secappa}
 
 In this appendix, we present various technical details which are used in the main part of the paper.

 \subsection{$\SO(8)$ Gamma matrices}
 \label{appa1}
 
 We are working with $8 \times 8$ Gamma matrices $\Gamma^I_{A \dot A}$ and their transposes $\Gamma^I_{\dot A A}$, which satisfy the Clifford algebra,
	\begin{align} \Gamma^I_{A \dot A} \Gamma^J_{\dot A B} + \Gamma^J_{A \dot A} \Gamma^I_{\dot A B} = 2 \delta^{IJ} \delta_{AB}
	\end{align}
Explicitly, we use the basis in \cite{Green:1987sp},
	\begin{align}
	\Gamma^8_{A \dot A} &= 1 \otimes 1 \otimes 1 &  \Gamma^1_{A \dot A} &= i \sigma_2 \otimes i \sigma_2 \otimes i \sigma_2  \nonumber \\
	\Gamma^2_{A \dot A} &= 1 \otimes \sigma_1 \otimes i\sigma_2 &  \Gamma^3_{A \dot A} &= 1 \otimes \sigma_3 \otimes i\sigma_2 \nonumber \\
	\Gamma^4_{A \dot A} &= \sigma_1 \otimes i\sigma_2 \otimes 1 &  \Gamma^5_{A \dot A} &= \sigma_3 \otimes i\sigma_2 \otimes 1  \nonumber \\
	\Gamma^6_{A \dot A} &= i\sigma_2 \otimes 1 \otimes \sigma_1 &  \Gamma^7_{A \dot A} &= i\sigma_2 \otimes 1 \otimes \sigma_3
	\end{align}
The matrices $\Gamma^{IJ}_{AB}$, $\Gamma^{IJ}_{\dot A \dot B}$ and similar are defined as unit-weight antisymmetrized products of Gamma matrices with the appropriate indices contracted.
For instance,
	\begin{align} \Gamma^{IJ}_{AB} \equiv \frac{1}{2} (\Gamma^I_{A \dot A} \Gamma^J_{\dot A B} - \Gamma^J_{A \dot A} \Gamma^I_{\dot A B}) 
	\end{align}
	
\subsection{Integrability conditions}
 \label{appa2}

For BPS equations of the form
	\begin{align*}
	\partial_t \epsilon &= - \frac{1}{2z} \gamma_0 \qty\big( \gamma_1 + f(u) + g(u) \gamma_2) \epsilon \nonumber \\
	\partial_z \epsilon &= - \frac{1}{2z} \gamma_1 \qty\big( f(u) + g(u) \gamma_2) \epsilon \nonumber \\
	\partial_u \epsilon &= \qty\big(F(u) + G(u) \gamma_2) \epsilon 
	\end{align*}
where $f, g, F, G$ are matrices acting on $\epsilon$ that commute with $\gamma_a$, the integrability conditions are 
	\begin{align*}
	t, z : \qquad 0 &=  (1 + f^2 + g^2)\epsilon + [f, g] \gamma_2 \epsilon \nonumber \\
	t, u : \qquad 0 &=  (f' + [f, F] - \{g, G\})\epsilon + (g' + [g, F] + \{f, G\}) \gamma_2 \epsilon \nonumber \\
	z, u: \qquad \phantom{0} & \qquad \text{same as for }t, u
	\end{align*}
	
\subsection{Scalar asymptotics}
 \label{appa3}

The asymptotic expansions of the $\phi_4$ and $\phi_5$ scalar fields, as given in (\ref{eq:scalar-sol}), in the limits $u \to \pm \infty$ are
\begin{align}
	|\phi_4| =&\; 2 |\sinh q| \frac{p \pm  q}{\sinh(p \pm q)} e^{\mp u} \nonumber \\
	&- \frac{2 |\sinh q|}{\sinh^2(p \pm q)} \qty( \frac{p \pm  q}{\sinh(p \pm q)} (\sinh^2 p + \sinh^2 q)  \pm 2 \sinh p \sinh q ) e^{\mp 3 u} + \cO(e^{\mp 5 u}) \nonumber \\
	\phi_5 =&\; (p \pm q) - \frac{2 }{\sinh(p \pm q)} \qty( \frac{p \pm q}{\sinh(p \pm q)} \sinh^2 q  \pm \sinh p \sinh q ) e^{\mp 2u} + \cO(e^{\mp 4u})
	\end{align}

 \newpage
\providecommand{\href}[2]{#2}\begingroup\raggedright\endgroup
\end{document}